\documentclass[prl,twocolumn,showpacs,amssymb,superscriptaddress]{revtex4}

\usepackage{graphicx}
\usepackage{epsfig}
\usepackage{dcolumn}
\usepackage{bm}

\newcommand{\comment}[1]{}
\newcommand{\newc}{\newcommand}

\def\thetab0{\theta_{B_0}}

\def\r2{\sqrt 2}
\def\beq{\begin{equation}}
\def\eeq{\end{equation}}
\def\bea{\begin{eqnarray}}
\def\eea{\end{eqnarray}}
\def\sinW2{\sin^2\theta_W}

\def\mz2{M_{z}^2}
\def\c2b{\cos 2\beta}

\def\mz{M_Z}

\def\sec2w{sec^2\theta_W}

\def\gmin2{(g-2)_\mu}

\catcode`\@=11 

\def\lsim{\mathrel{\mathpalette\@versim<}}
\def\gsim{\mathrel{\mathpalette\@versim>}}
\def\@versim#1#2{\vcenter{\offinterlineskip
    \ialign{$\m@th#1\hfil##\hfil$\crcr#2\crcr\sim\crcr } }}

\def\mpT{p_T \hspace{-1em}/\;\:}

\newc{\ra}{\rightarrow}
\newc{\s}{\smallskip}
\newc{\nn}{\noindent}
\newc{\non}{\nonumber}

\def \chonep{{\wt\chi_1}^{+}}
\def \chonem{{\wt\chi_1^-}}
\def \chonep2{{\wt\chi_2^+}}
\def \chonem2{{\wt\chi_2^-}}

\begin{document}
\begin{flushright}
\hspace*{6in}
\mbox{HIP-2008-33/TH}
\end{flushright}
%
%
\title{Correlations between sneutrino oscillations and signatures at the LHC in anomaly-mediated supersymmetry breaking}

\author{Dilip Kumar Ghosh}%
\email{dghosh@prl.res.in}
\affiliation{Theoretical Physics Division, Physical Research Laboratory, 
Navrangpura, Ahmedabad 380 009, India\footnote{After 26 February 2009, permanent institute is $^3$.}} 
\author{Tuomas Honkavaara}%
\email{Tuomas.Honkavaara@helsinki.fi}
\affiliation{Department of Physics, and Helsinki Institute of Physics,
P.O.Box 64, FIN-00014 University of Helsinki, Finland} 
\author{Katri Huitu}%
\email{Katri.Huitu@helsinki.fi}
\affiliation{Department of Physics, and Helsinki Institute of Physics,
P.O.Box 64, FIN-00014 University of Helsinki, Finland}
\author{Sourov Roy}%
\email{tpsr@iacs.res.in}
\affiliation{Department of Theoretical Physics and Centre for Theoretical
Sciences, Indian Association for the Cultivation of Science, 2A $\&$ 2B Raja
S.C. Mullick Road, Kolkata 700 032, India}
\date{\today}
\begin{abstract}
Sneutrino-antisneutrino oscillation can be observed at the LHC by studying a charge asymmetry of the leptons in the final states. We demonstrate this in the context of an anomaly-mediated supersymmetry breaking model which can give rise to a large oscillation probability. The preferred region of the parameter space is characterized by the presence of a sneutrino next-to-lightest supersymmetric particle and a stau lightest supersymmetric particle. We show that the signals studied here have certain correlations with the pattern of the sneutrino oscillation.
\end{abstract}



\maketitle

Sneutrino-antisneutrino mixing occurs in any supersymmetric (SUSY) model where neutrinos have nonzero Majorana masses. Such $\Delta L$ = 2 Majorana 
neutrino mass terms can induce a mass splitting ($\Delta m_{\tilde \nu}$) 
between the physical states. 
The effect of this mass splitting is to induce sneutrino-antisneutrino 
oscillations  \cite{hirschetal, grossman-haber1}. This can lead to the sneutrino decaying into a final state with a ``wrong-sign charged lepton," and the lepton number can be tagged in sneutrino decays by the charge of the final state lepton. Here, we assume that the sneutrino flavor oscillation is absent and lepton flavor is conserved in the decay of sneutrino/antisneutrino.

As discussed in \cite{tuomas}, the probability
of finding a wrong-sign charged lepton in the decay of a sneutrino should 
be the time-integrated one and is given by
\bea
P(\tilde \nu \rightarrow \ell^+) = {\frac {x^2_{\tilde \nu}}
{2(1+x^2_{\tilde \nu})}}  {\cal B}_{\tilde \nu^{\ast}}({\tilde \nu}^{\ast}
 \rightarrow \ell^+ X)
\label{eqn-prob},
\eea
where the quantity $x_{\tilde \nu}$ is defined as
$ x_{\tilde \nu} \equiv \Delta m_{\tilde \nu}/\Gamma_{\tilde \nu} $,
and ${\cal B}_{{\tilde \nu}^{\ast}}$ is the branching ratio for 
$\tilde \nu^* \rightarrow \ell^+$.
This signal can be observed from the single production of a sneutrino at 
the LHC, provided $ x_{\tilde \nu} \sim 1$ 
and ${\cal B}_{{\tilde \nu}^{\ast}}$ is significant.

In this paper, we demonstrate that a particular SUSY mass spectrum
possible within the framework of an anomaly-mediated supersymmetry breaking
model (AMSB) \cite{amsb} can lead to such a signature at the LHC.
It is evident from the above discussion that the probability of the 
sneutrino-antisneutrino oscillation depends crucially on 
$\Delta m_{\tilde \nu}$ and $\Gamma_{\tilde \nu}$. 
If $m_\nu \sim 0.1$ eV, the radiative corrections to the 
$m_\nu$ induced by $\Delta m_{\tilde \nu}$ face the bound
\cite{grossman-haber1}
$\Delta m_{\tilde \nu}/m_\nu \lsim \mathcal O (4\pi/\alpha)$, 
implying $\Delta m_{\tilde \nu} \lsim 0.1$ keV. 
Thus, in order to get $x_{\tilde \nu} \sim 1$, one also needs the sneutrino decay width $\Gamma_{\tilde \nu}$ to be $\sim \Delta m_{\tilde \nu} $. Because of the smallness of $\Gamma_{\tilde \nu}$, the sneutrino's lifetime would be large enough for sneutrino oscillation to take place before its decay. 

However, for a spectrum where ${\tilde \chi^0_1}$ is the lightest supersymmetric particle (LSP), $\Gamma_{\tilde \nu} $ would not be $\lsim {\cal O}(1)$ keV because of the presence of two-body decays:
$\tilde \nu \rightarrow \nu \tilde\chi^0$ and $\tilde \nu \rightarrow \ell^- \tilde \chi^+$. If, instead, the mass spectrum is such that
\bea
m_{{\tilde \tau}_1} < m_{\tilde \nu} < m_{\tilde \chi^0_1}, m_{\tilde
\chi^\pm_1},
\label{spectrum}
\eea
where the lighter stau (${\tilde \tau}_1$) is the LSP,
these two-body
decay modes are forbidden and 
the three-body decay modes such as $\tilde \nu \rightarrow 
\ell^- {\tilde \tau}_1^+ \nu_\tau$ and $\tilde \nu \rightarrow \nu 
{\tilde \tau}_1^\pm \tau^\mp$ are the available ones. In addition,
the branching fraction of $\tilde{\nu}^* \rightarrow \ell^+ {\tilde \tau}_1^- \bar{\nu}_\tau$ final state gives the wrong-sign charged lepton signal. 
However, having
${\tilde \tau}_1$ as a stable charged particle is strongly disfavored by
astrophysical grounds. This can be avoided, for example,
if a very small $R$-parity-violating coupling $ (\lsim 10^{-8})$ 
induces the decay ${\tilde \tau}_1 \rightarrow \ell \nu$, which
occurs outside the detector after producing a heavily
ionized charged track in the detector. 

\begin{figure}\label{parameter_space}
\centering
\includegraphics[height=6.00cm]{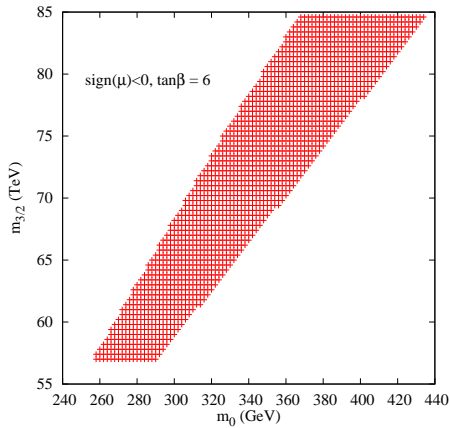}
\caption{Parameter space of the AMSB model with sneutrino NLSP and 
${\tilde \tau}_1$ LSP and $m_{\tilde{\ell}_{1,2}}<m_{\tilde{\chi}_1^0,\tilde{\chi}_1^\pm}$ ($\ell=e,\mu$).}
\label{amsb-para-space}
\end{figure}

The required spectrum (\ref{spectrum}) 
can be obtained in some region of the AMSB parameter space 
with $\Delta m_{\tilde \nu} \lsim \mathcal O (4\pi m_\nu/\alpha)$. In our analysis, we have $m_{\nu_i} \lsim 0.3$ eV ($i=e,\mu,\tau$). In Fig. \ref{amsb-para-space}, we display the region of the parameter space in $m_0 - m_{3/2}$ plane with ${\rm sign}(\mu) < 0 $ and $\tan\beta = 6$, where the above spectrum is valid. It is worthwhile to mention that the three sneutrinos are almost mass degenerate (${\tilde \nu}_\tau$ is slightly lighter though) and they are the NLSP. In this parameter space, $\tilde \chi^0_1$ and $\tilde \chi^{\pm}_1$ are heavier than the other charged sleptons.

\begin{figure}\label{osc-prob}
\centering
\includegraphics[height=6.50cm]{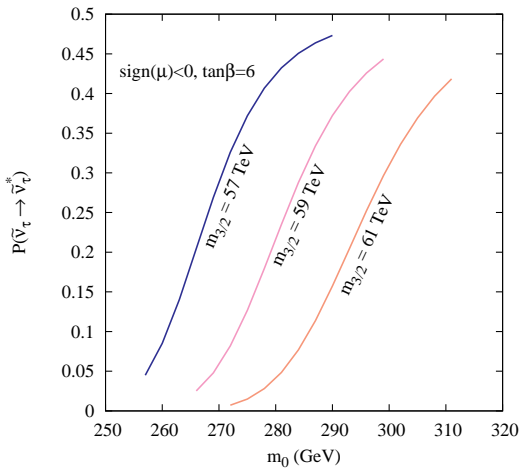}
\caption{${\tilde \nu}_\tau$ oscillation probability as a function of
$m_0$.}
\label{osc-prob-tanb-7}
\end{figure}

In Fig. \ref{osc-prob-tanb-7}, we plot the ${\tilde \nu}_\tau$ oscillation 
probability as a function of the common scalar mass $m_0$ for three different choices of $m_{3/2}$ with ${\rm sign}(\mu) < 0 $ and 
$\tan\beta = 6$ in the allowed parameter space. 
It can be seen from this figure that the probability
of oscillation can go as high as 0.45, and it is this probability that we 
propose to measure. Hence, the AMSB has a good potential to 
produce signals of sneutrino-antisneutrino oscillation, which can be tested in colliders. This has been noticed earlier; see \cite{like-sign-norp,tuomas}.

Let us remark that the signatures of 
sneutrino oscillation at the future $e^+e^-$ and $e^- \gamma$ linear colliders have been studied in Refs. \cite{like-sign-norp, tuomas}, whereas authors in Ref. \cite{dedes} discussed sneutrino oscillation in the seesaw extended minimal supersymmetric standard model (MSSM).  
In this work, the first production process we will consider is  
\begin{eqnarray}
p p  \to {\tilde \nu}_\tau {\tilde \tau}_1^+
\label{ppstau1}.
\end{eqnarray}
Since ${\tilde \nu}_\tau$ decaying to a three-body final state with $\tau^\pm$ 
is difficult to identify due to the small branching ratio and 
$\tau$ detection efficiency, we look at other channels mediated by 
virtual $W^-$ and $H^-$.
If the 
${\tilde \nu}_\tau$ oscillates into a ${\tilde \nu}_\tau^*$, we can have 
a three-body final state, ${\tilde \nu}_\tau \rightarrow {\tilde \nu}_\tau^* \rightarrow \ell^- {\tilde \tau}_1^+ {\bar \nu}_{\ell}$ leading 
to $\ell^- {\tilde \tau}_1^+ {\tilde \tau}_1^+ + \mpT \,$ signature from 
the process in Eq. (\ref{ppstau1}). 
Here, $\ell= e,~\mu$.
The cross section for this process 
is given by $
\sigma^1_\mathrm{osc}= \sigma(p p  \rightarrow {\tilde \nu}_\tau 
{\tilde \tau}_1^+) \times P_{{\tilde \nu}_\tau \rightarrow 
{\tilde \nu}_\tau^*} \times {\cal B}_{\tilde \nu^\ast}
({\tilde \nu}_\tau^* \rightarrow
\ell^- {\tilde \tau}_1^+ {\bar \nu}_{\ell}) $, where 
$P_{{\tilde \nu}_\tau \rightarrow {\tilde \nu}_\tau^\ast}$ denotes the 
sneutrino oscillation probability.
When the ${\tilde \nu}_\tau$ survives as a ${\tilde \nu}_\tau$, one of the possible three-body decays of the ${\tilde \nu}_\tau$ is ${\tilde \nu}_\tau \rightarrow \ell^+ {\tilde \tau}_1^- \nu_{\ell}$. 
This would lead to $\ell^+ {\tilde \tau}_1^- {\tilde \tau}_1^+ + \mpT $~ 
signature
from the same process (\ref{ppstau1}). 
Similar to the oscillation case, the cross section for this process is given by 
$\sigma^1_\mathrm{no \ osc} = \sigma(p p  \rightarrow {\tilde \nu}_\tau {\tilde \tau}_1^+) \times P_{{\tilde \nu}_\tau \rightarrow 
{\tilde \nu}_\tau} \times {\cal B}_{\tilde \nu} ({\tilde \nu}_\tau \rightarrow
\ell^+ {\tilde \tau}_1^- {\nu_\ell}) $, where 
$P_{{\tilde \nu}_\tau \rightarrow {\tilde \nu}_\tau }$ denotes the 
sneutrino survival probability.

As mentioned earlier, in the presence of 
a very small $R$-parity violating coupling, both of these staus would 
decay outside the detector after traversing the whole length of the detector producing heavily ionized charged track. In this case, the lepton number can be tagged by the charge of the long-lived stau.

Let us define a charge asymmetry parameter in terms of the
signal from sneutrino oscillation and no-oscillation:
\begin{eqnarray}
A_\mathrm{asym} \equiv {{\sigma (\ell^-\tilde\tau_1^+\tilde\tau_1^+ +\mpT)
-\sigma (\ell^+\tilde\tau_1^-\tilde\tau_1^+ +\mpT)}\over
{\sigma (\ell^-\tilde\tau_1^+\tilde\tau_1^+ +\mpT)
+\sigma (\ell^+\tilde\tau_1^-\tilde\tau_1^+ +\mpT)}}
\label{asymosc}
\end{eqnarray}
Since ${\cal B}_{\tilde \nu}({\tilde \nu}_\tau 
\rightarrow \ell^+ {\tilde \tau}_1^- 
{\nu_\ell}) = {\cal B}_{\tilde \nu^\ast} 
({\tilde \nu}_\tau^* \rightarrow \ell^- {\tilde \tau}_1^+ 
{\bar \nu}_{\ell}) $, one can rewrite Eq. (\ref{asymosc}) 
in the following form 
\begin{eqnarray}
A_\mathrm{asym} = P_{{\tilde \nu}_\tau \rightarrow {\tilde \nu}_\tau^*} 
-P_{{\tilde \nu}_\tau \rightarrow {\tilde \nu}_\tau}.
\label{asym1}
\end{eqnarray}

It is evident from Eq. (\ref{asym1}) that $A_{asym} 
= -1$ corresponds to no sneutrino oscillation. Hence, any deviation 
of $A_{asym}$ from -1 is the smoking gun signature of sneutrino 
oscillation. From Eq. (\ref{asym1}),
it is clear that the measurement of the asymmetry immediately tells us about 
the sneutrino-antisneutrino oscillation probability $P_{{\tilde \nu}_\tau 
\rightarrow {\tilde \nu}_\tau^*}$ which is a function of the ratio 
$x_{\tilde \nu} \equiv \Delta m_{\tilde \nu}/\Gamma_{\tilde \nu}$. Hence, once
the total sneutrino decay width is known, one can easily calculate the 
mass splitting between the sneutrino and the antisneutrino. This 
mass splitting is proportional to the neutrino mass, and, thus, the 
measurement of the asymmetry gives us the absolute value of 
the neutrino mass. This is an alternative way to the neutrinoless 
double beta decay experiments to probe the absolute value of the 
neutrino mass. Let us note that there is very little SM background 
to these signals assuming that the long-lived staus produce heavily 
ionized charged tracks which can be distinguished from the muon 
tracks. This is possible, since the staus are much slower than
the muons because of their large masses.

However, there are several other SUSY processes which can give rise 
to the same final state as our signal. Part of these processes has 
a sneutrino in the final state  and the other part is without a sneutrino. 
These processes are
\bea
&&pp\rightarrow \tilde\nu_\ell \tilde \ell_L^+ \,\, {\rm with}\,\, \ell=e,\mu ,
\quad {\rm and}\quad
pp\rightarrow \tilde\chi_1^0\tilde\chi_1^+.
\eea
Here, the relevant decay modes, which can lead to the same signal as in 
Eq. (\ref{ppstau1}), are for
${\tilde \chi}_1^0$, ${\tilde \chi}^+_1$, and ${\tilde \ell}^+_L$: \\ 
$ {\tilde \chi}_1^0 \to {\tilde \nu}_\ell \bar{\nu}_\ell, 
{\tilde \nu}_\ell^* \nu_\ell, {\tilde \nu}_\tau {\bar \nu}_\tau,  
{\tilde \nu}_\tau^* \nu_\tau, 
{\tilde \chi}^+_1 \to {\tilde \ell}^+_L \nu_\ell, 
{\tilde \tau}^+_1 \nu_\tau,
{\tilde \ell}^+_L \to {\tilde \tau}^+_1 \nu_\tau \bar{\nu}_\ell $,
where $\ell = e $ or $\mu $.
Since the charged sleptons are lighter than the neutralinos and
the charginos, they decay to three bodies.
If ${\tilde \nu}_\ell$ oscillates, it will decay to a wrong-sign charged 
lepton. The relevant ${\tilde \nu}_\ell$ decays are 
$ {\tilde \nu}_\ell \to \ell^+ {\tilde \tau}^-_1 {\bar \nu}_\tau 
~~~({\rm with ~oscillation}) $ and 
$ {\tilde \nu}_\ell \to \ell^- {\tilde \tau}^+_1 \nu_\tau 
~~~({\rm without ~oscillation}) $
and correspondingly for ${\tilde \nu}^*_\ell$.
The 
relevant decays of ${\tilde \nu}_\tau$ and ${\tilde \nu}_\tau^*$ have been
discussed earlier.

Thus, for example, $pp \to
{\tilde \chi}_1^0 {\tilde \chi}^+_1 \to ({\tilde \nu}_\ell \bar{\nu}_\ell) 
({\tilde \ell}^+_L \nu_\ell)$ can give rise to the signal $pp \to
\ell^- {\tilde \tau}^+_1 {\tilde \tau}^+_1 + \mpT$ without ${\tilde \nu}_\ell$
oscillation, whereas the same decay chain can also produce $pp \to \ell^+ 
{\tilde \tau}^-_1 {\tilde \tau}^+_1 + \mpT$ for oscillating
${\tilde \nu}_\ell$. Similarly, $pp \to {\tilde \nu}_\ell 
{\tilde \ell}^+_L$ leads to $\ell^+ {\tilde \tau}^-_1 {\tilde \tau}^+_1 + 
\mpT$ with the oscillation of the ${\tilde \nu}_\ell$ into a 
${\tilde \nu}_\ell^*$. 
The same production process gives
rise to the final state $\ell^- {\tilde \tau}^+_1 {\tilde \tau}^+_1 + \mpT$ when ${\tilde \nu}_\ell$ does not oscillate. $\tilde{\chi}_1^0$ can also 
decay to a tau-sneutrino and a tau-neutrino, leading to the final states 
$pp \to  \ell^+ {\tilde \tau}^-_1 {\tilde \tau}^+_1 + \mpT$ and $pp \to \ell^- 
{\tilde \tau}^+_1 {\tilde \tau}^+_1 + \mpT$. Heavier neutralinos and chargino 
are not considered in the production process because of their much smaller
cross sections. 

The signal can also result from $\tilde\chi_1^0$ cascade decay without
any intermediate sneutrinos.
For example, $pp \to {\tilde \chi}_1^0 
{\tilde \chi}^+_1 \to (\tau^-\tilde{\tau}_1^+)(\tilde{\tau}_1^+ \nu_\tau)$
can lead to $\ell^- {\tilde \tau}^+_1 {\tilde \tau}^+_1 + \mpT$, where 
$\tau^-$ decays leptonically. 
Similarly, $pp \to {\tilde \chi}_1^0 {\tilde \chi}^+_1 
\to (\ell^+ \tilde{\ell}_L^-)(\tilde{\tau}_1^+ \nu_\tau)$ can produce 
$\ell^+ {\tilde \tau}^-_1 {\tilde \tau}^+_1 + \mpT$.   

All the processes discussed above can be considered as the SUSY backgrounds
to the signals originating from the process (\ref{ppstau1}). 
The charge asymmetry $A_{asym}$ must be considered
after taking into account these backgrounds in order to 
observe the signature of ${\tilde \nu}_\tau$ oscillation. One should note 
that processes involving ${\tilde \nu}_\ell$/${\tilde \nu}_\ell^*$ 
contribute to the final states $\ell^+ {\tilde \tau}^-_1 {\tilde \tau}^+_1 
+ \mpT$ and $\ell^- {\tilde \tau}^+_1 {\tilde \tau}^+_1 + \mpT$ in an 
opposite manner compared to the signal processes. 
Hence, it is indispensable to find suitable 
kinematical cuts to remove these backgrounds so that the signatures of 
${\tilde \nu}_\tau$ oscillation can be observed through the measurements 
of the lepton charge asymmetry defined in Eq. (\ref{asymosc}).

Nevertheless, we observe that the 
distributions of two kinematical quantities, namely, the transverse momentum ($p_T$) of the staus and the missing transverse momentum ($\mpT$) are very crucial in distinguishing these SUSY background processes from the signals. It turns out that, using suitable cuts, one can have $S/\sqrt{B} \approx 5$ or higher even with an integrated luminosity of 30 fb$^{-1}$.    

We select the signal events with the following criteria$\colon$ 
1) $p^{\ell^\pm}_T> 5$ GeV, 2) $|\eta^{{\ell^\pm},{\tilde \tau}_1}|< 2.5$,
3) transverse momentum of both ${\tilde \tau}_1^-$'s must satisfy 
$p^{{\tilde \tau}_1}_T> 100$ GeV, and 4) $\mpT < 20$ GeV. The last mentioned
two cuts are crucial in clearly identifying signals from the SUSY 
background. 

In Table \ref{table1}, we listed the cross sections of 
final states described above originating from different 2 $\to$ 2 production processes for a particular parameter point, namely, $\tan\beta$ = 6, $\mu <$ 0, and $m_0$ = 270 GeV, and $m_{3/2}$ = 57 TeV. These numbers are obtained after applying the selection cuts. One can see that the SUSY background events are well suppressed compared to the signal events. Now, if we calculate the asymmetry, as defined in Eq. (\ref{asymosc}), from the numbers in the first row of Table \ref{table1}, the asymmetry comes out to be  -0.42. If we solve from this for the ${\tilde \nu}_\tau$ oscillation 
probability using Eq. (\ref{asym1}), we get $P_{{\tilde \nu}_\tau \rightarrow {\tilde \nu}_\tau^*}$ = 0.29, which can be compared with Fig. (\ref{osc-prob-tanb-7}). 
\begin{table}
\begin{center}
\footnotesize
\begin{tabular}{|c|c|c|c|c|}
\hline
{} &
\multicolumn{4}{c|}{Cross sections in fb} \\
\cline{2-5}
Process & $e^- {\tilde \tau}^+_1 {\tilde \tau}^+_1 \mpT$  &
$\mu^- {\tilde \tau}^+_1 {\tilde \tau}^+_1 \mpT$ & 
$\mu^+ {\tilde \tau}^-_1 {\tilde \tau}^+_1 \mpT$ &
$e^+ {\tilde \tau}^-_1 {\tilde \tau}^+_1 \mpT$ 
\\
\hline
$pp \to {\tilde \nu}_\tau {\tilde \tau}_1^+$ & 1.259 & 1.259 & 3.095 & 
3.095
\\
\hline
$pp \to {\tilde \nu}_\ell {\tilde \ell}^+_L$ & 0.373 & 0.373 & 0.303 & 
0.303
\\
\hline
$pp \rightarrow {\tilde \chi}_1^0 {\tilde \chi}^+_1$ & 0.206 & 0.206 & 
0.206 & 0.206
\\
(with $\tilde{\nu}$'s) & & & &
\\
\hline
$pp \rightarrow {\tilde \chi}_1^0 {\tilde \chi}^+_1$ & 0.036 & 0.036 & 
0.036 & 0.036
\\
(no $\tilde{\nu}$'s) & & & &
\\
\hline
\end{tabular}
\end{center}
\caption{Cross sections for $pp \to \ell^- {\tilde \tau}^+_1 
{\tilde \tau}^+_1 + \mpT$ and $pp \to \ell^+ {\tilde \tau}^-_1 
{\tilde \tau}^+_1 + \mpT$ from different two-body production processes.
Here, $\tan\beta$ = 6, $\mu <$ 0,
$m_0$ = 270 GeV, and $m_{3/2}$ = 57 TeV.} 
\label{table1}
\end{table}
However, as discussed, the SUSY background leads to the same final
states and has to be taken into account when calculating the asymmetry.

It should be kept in mind though that, in this situation, there are two different oscillation probabilities involved. One is the tau-sneutrino oscillation probability and the other is the electron (muon) sneutrino oscillation probability. Hence, by measuring the asymmetry as defined in Eq. (\ref{asymosc}), it would not be possible to measure these two oscillation probabilities. In any case, the measurement of the asymmetry can clearly indicate whether the sneutrinos are oscillating or not. For example, let us again look at the parameter point studied in Table \ref{table1} and assume that the SUSY background contamination is not possible to be better separated from the signal. In such a situation, the resultant asymmetry is -0.33 when the sneutrinos are oscillating and -0.68 when there is no sneutrino oscillation. Compared to the ideal case with no SUSY background, the measurable asymmetry in the sneutrino oscillation case changes by less than 15 \%.

In Table \ref{table2}, 
we show the asymmetries including the SUSY background for three
different parameter choices. 
In all of these cases, the oscillation probability is more than $0.15$.
If the parameter set is already known from the measurements of the SUSY
spectrum, the expectation for the asymmetry is also known.  
It is seen from the table that, already with 30 fb$^{-1}$, one can 
distinguish between the oscillation and no-oscillation cases in these
sample points.
When $\tan\beta$ grows, the ratio between the SUSY signal and the background reduces. Thus, this measurement, with the cuts used, is possible for small $\tan\beta$. We find relevant parameter sets for $\tan\beta=5,6,7$, which have been shown in Table \ref{table2}.

\begin{figure}
\centering
\includegraphics[height=6.00cm]{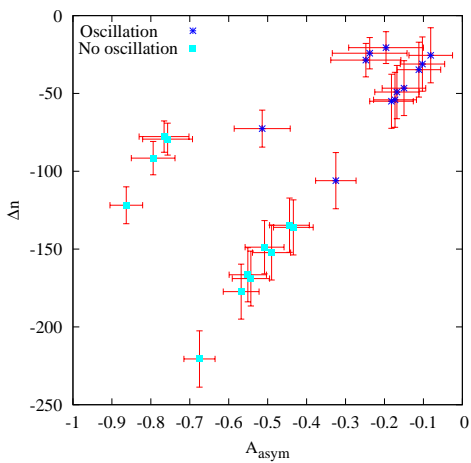}
\caption{Correlation between $\Delta n$ and $A_\mathrm{asym}$ for the
oscillation and the no-oscillation cases for different parameter points.}
\label{correlation}
\end{figure}

If the SUSY spectrum is not known, one can still deduce in favorable
cases whether there is sneutrino oscillation or not. We demonstrate this
in Fig.~\ref{correlation} for $\tan\beta=5,6$ and the 
values of $m_0$ and $m_{3/2}$ for which the signal 
cross sections are large. Here, it has been required that oscillation probability is more than
0.25 and $S/\sqrt{B} \gsim 5 $.
We plot
the difference ($\Delta n$) between the numbers of events for 
$pp \to \ell^- {\tilde \tau}_1^+ {\tilde \tau}_1^+ + \mpT$ and 
$pp \to \ell^+ {\tilde \tau}_1^- {\tilde \tau}_1^+ + \mpT$ 
for integrated luminosity 30 fb$^{-1}$ vs the asymmetry. The corresponding
errors are shown at the $1 \sigma$ level.
\begin{table}
\begin{center}
\footnotesize
\begin{tabular}{|l|c|c|c|c|}
\hline
Parameter point & $A_\mathrm{asym}$ & 
\multicolumn{3}{c|}{$\pm$ Errors} \\
\cline{3-5}
$\tan\beta$, $m_0$(GeV), &osc. &  &  & \\
$m_{3/2}$(TeV) &(no osc.)& 30 fb$^{-1}$ & 100 fb$^{-1}$& 300 fb$^{-1}$ \\
\hline
5, 370, 81, $\mu < 0 $ & -0.515 & 0.072 & 0.040 &0.023 \\
& (-0.859) & (0.043) & (0.024) & (0.014)
\\
\hline
6, 270, 57, $\mu < 0 $ & -0.325 & 0.052 & 0.029 & 0.017 \\
& (-0.676) & (0.041) & (0.022) & (0.013)
\\
\hline
7, 248, 49, $\mu < 0 $ & -0.149 & 0.044 & 0.024 & 0.014 \\
& (-0.266) & (0.043) & (0.024) & (0.014)
\\
\hline
\end{tabular}
\end{center}
\caption{Asymmetries and the corresponding errors for different parameter
points. Numbers in the brackets are for the no-oscillation case.} 
\label{table2}
\end{table}
%
One can see from this correlation plot that the sneutrino oscillation 
represents bigger asymmetry and bigger $\Delta n$, whereas, in the
case of no sneutrino oscillation, the value of $\Delta n$ and the 
asymmetry should be on the smaller side. This is expected, since, with the cuts that we have imposed, the ${\tilde \nu}_\tau$ type of oscillation signal is stronger.
When there 
is oscillation, the splitting between two different charge final states is
smaller, and, naturally, the asymmetry is closer to zero.

It is interesting to note that one can also make the processes involving the 
${\tilde \nu}_{e/\mu}$ dominate over the process $pp \to {\tilde \nu}_\tau 
{\tilde \tau}_1^+$ by switching the missing $p_T$ cut, i.e., by using
$\mpT >$ 20 GeV. In this case, the asymmetry comes out with an opposite sign.
By looking at the sign of the asymmetry parameter and with 
appropriate missing $p_T$ cut, one can conclude whether the processes involve 
predominantly ${\tilde \nu}_\tau$ or ${\tilde \nu}_{e/\mu}$. This is another 
remarkable feature of this study. However, in this case, one gets, 
in general, a smaller asymmetry because of larger total cross 
section (which makes the errors smaller though). For $\tan\beta=5,6$, the 
${\tilde \nu}_{e/\mu}$ oscillation probability is close to 0.5.  
Hence, one would expect close to zero asymmetry for such a parameter 
point. If there is no oscillation, then the asymmetry is larger, but it never goes to $+1$ because of the contributions from $\tilde{\chi}_1^0\tilde{\chi}_1^+$ production. 

It is important that, in this study, we have assumed that the staus
decay outside the detector. It is also possible that the $R$-parity violating coupling is larger and the staus decay inside the detector after traversing a certain length or they decay promptly. Obviously, the signals discussed here will be changed in those situations, and the analysis to find out the asymmetry will also be different. A detailed discussion on these issues is beyond the scope of this paper, and we hope to come back to these in a future work \cite{sneutrino_next_work} along with some more details of the present study. One could also look at the production process $p p  \rightarrow 
{\tilde \nu}_\tau^* {\tilde \tau}_1^-$.  However, the cross section is smaller 
than in Eq. (\ref{ppstau1}) \cite{beenakker}.  

In conclusion, we have demonstrated that a suitably defined lepton charge 
asymmetry may provide us information about sneutrino oscillation at the LHC. 
This scenario can be realized in an AMSB model with Majorana neutrino masses. 
Although distinguishing between sneutrino oscillation and no-oscillation is
much easier when the SUSY spectrum is known, we have shown that it is also
possible to find out the oscillating sneutrino even without the knowledge of 
the spectrum by looking at simple correlations. One can also provide 
information about the absolute neutrino mass scale from this study when the 
SUSY spectrum is known. 

We are grateful to D. Choudhury, R. Kinnunen, and S.K. Rai 
for helpful discussions. 
This work is supported in part by the Academy of Finland 
(Project No. 115032).
DKG and SR acknowledge the hospitality provided by the Helsinki Institute of
Physics. 
TH thanks the V\"ais\"al\"a Foundation for support.

\end{document}